\documentclass[twocolumn, prx, superscriptaddress, notitlepage]{revtex4-1}
\usepackage{graphicx}
\usepackage{physics}
\usepackage{float} 
\usepackage{subfigure} 
\usepackage{color}
\usepackage{amsmath}
\usepackage{amsfonts}
\usepackage[normalem]{ulem}
\usepackage{xspace}
\usepackage[dvipsnames]{xcolor}
\usepackage{dcolumn}
\usepackage[breaklinks,colorlinks,linkcolor=blue,citecolor=blue,urlcolor=blue]{hyperref}

\begin{document}
\title{Hidden curved spaces in Bosonic Kitaev model}

\author{Chenwei Lv}
\affiliation{Department of Physics and Astronomy, Purdue University, West Lafayette, IN, 47907, USA}
\affiliation{Homer L. Dodge Department of Physics and Astronomy, University of Oklahoma, Norman, OK, 73019, USA}

\author{Qi Zhou}
\email{zhou753@purdue.edu}
\affiliation{Department of Physics and Astronomy, Purdue University, West Lafayette, IN, 47907, USA}
\affiliation{Purdue Quantum Science and Engineering Institute, Purdue University, West Lafayette, IN 47907, USA}

\date{\today}

\begin{abstract}
Quantum matter in curved spaces exhibits remarkable properties unattainable in flat spaces. To access curved spaces in laboratories, the conventional wisdom is that physical distortions need to be implemented into a system. In contrast to this belief, here, we show that two hyperbolic surfaces readily exist in bosonic Kitaev model in the absence of any physical distortions and give rise to a range of intriguing phenomena, such as chiral quantum transport or chiral reaction-diffusion. A finite chemical potential couples these two hyperbolic surfaces, delivering a quantum sensor whose sensitivity grows exponentially with the size of the system. Our results provide experimentalists with an unprecedented opportunity to explore intriguing quantum phenomena in curve spaces without distortion or access non-Hermitian phenomena without dissipation. Our work also suggests a new class of quantum sensors in which geometry amplifies small signals.  

\end{abstract}
\maketitle 

Whereas quantum effects in curved spaces produce the most fascinating phenomena in nature~\cite{Hawkingradiation,Unruh,Unruh0,Wald1995,Mukhanov2007}, it is challenging to detect the interplay between quantum mechanics and curved spaces in observational cosmology. 
As such, physicists have started to explore a variety of quantum systems in which the underlying metric could be controlled in laboratories. 
For instance, acoustic black holes and tunable curvatures have been realized for phonons in Bose-Einstein condensates~\cite{Unruh1995,Zoller2000,Steinhauer2010,Steinhauer2014, Steinhauer2016, MuozdeNova2019, Kolobov2021, Campbell2018, Campbell2022, Floerchinger2022, Viermann2022}. 
In parallel, superconducting and electric circuits may be used to realize certain tiling of hyperbolic surfaces~\cite{Kollr2019,Zhang2022,Lenggenhager2022,Chen2023}, and photonic devices have been engineered to study general relativity~\cite{Philbin2008, Genov2009, Leonhardt2009, Belgiorno2010, Sheng2013, Bekenstein2015, Bekenstein2017, Drori2019}.
These synthetic curved spaces have shed new light onto longstanding questions in quantum field theories and also provided physicists with a powerful tool to discover new quantum matter unattainable in flat spaces. 

Despite that the aforementioned efforts focus on very different atomic, electronic, and photonic platforms, they share a common feature: certain physical distortions have been applied. 
The density or the interaction was made non-uniform in a condensate, the superconducting circuits were distributed non-uniformly, and the refractive index changed in space. 
This is consistent with the generic belief that to make a curved manifold out of a flat space, certain physical distortions are required. 
Here, we point out that this conventional wisdom may not always be true. 
A quantum system could intrinsically embed curved spaces without any physical distortions. 
As such, extraordinary quantum phenomena emerging from such a quantum system have deep geometric roots. 
Our finding thus offers physicists a completely new routine to explore synthetic curved spaces. 

\begin{figure}
    \includegraphics[width=0.495\textwidth]{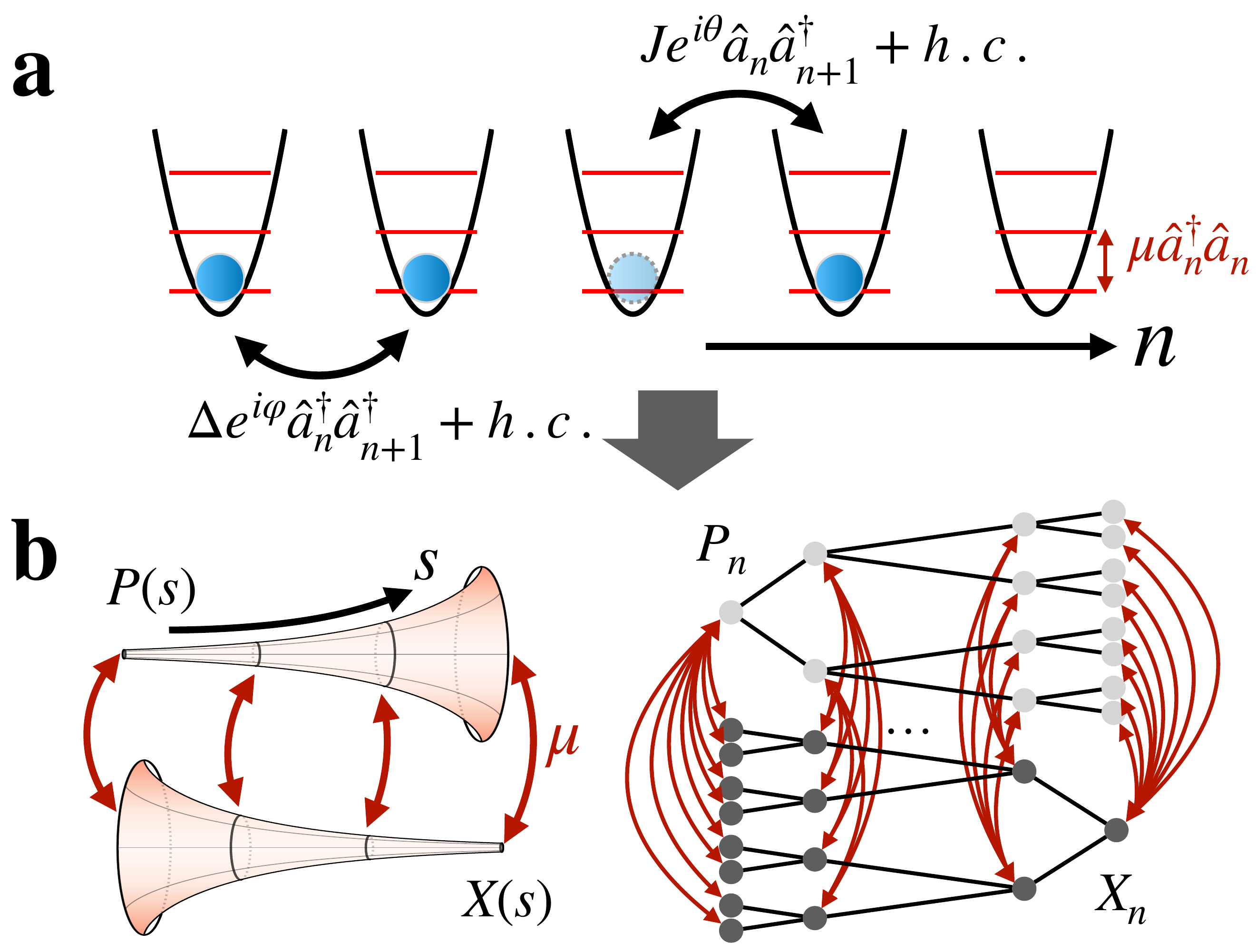}
    \caption{
    (a) A schematic of bosonic Kitaev model with the tunneling term $Je^{i\theta}$ and the pair creation term $\Delta e^{i\varphi}$. 
    (b) A one-dimensional bosonic Kitaev model is a dimension reduction of two hyperbolic spaces.
    The dimension reduction is obtained by choosing the periodic boundary condition in the azimuthal direction of the hyperbolic surfaces. 
    When $\mu$ vanishes, these two hyperbolic surfaces are decoupled. 
    The finite curvature forces the excitation in $P$ and $X$ to propagate towards opposite directions. 
    }\label{fig:Kitaev}
\end{figure}

To be explicit, we find that bosonic Kitaev model~\cite{Clerk2018}, a generalization of the celebrated fermionic Kitaev model~\cite{Kitaev2001}, encloses two underlying hyperbolic surfaces. 
Bosonic Kitaev model could also be considered as the large spin limit of the the spin Kitaev model~\cite{Zhou2011}. 
We find that the pair creation/annihilation terms in bosonic Kitaev model supply finite curvatures to the system, as depicted in Fig.~\ref{fig:Kitaev}. 
Each of these two hyperbolic surfaces can be visualized using a pseudosphere of a funnel shape, and the funneling mouths are located at the two opposite edges of the system. 
These two hyperbolic surfaces provide the source of a variety of prominent phenomena discussed in this work. 
When $\mu=0$, these two hyperbolic surfaces are decoupled. 
The excitation in one hyperbolic surface propagates towards a unique direction set by the location of the funneling mouth, while the excitation in the other hyperbolic surface travels towards the opposite direction, resulting in the chiral skin effect. 
A finite $\mu$ couples these two hyperbolic surfaces and the system becomes extremely sensitive to a small value of $\mu$, as the finite curvatures exponentially enhance the effect of $\mu$. 
Our findings unfold the geometric origin of the critical skin effect and also suggest a geometry-induced supersensitive quantum sensor. 

We consider a one-dimensional lattice model, 
\begin{equation}
  \hat H_b=\sum_n \mu \hat a_n^\dag \hat a_n + \begin{pmatrix}\hat a_n & \hat a_n^\dag \end{pmatrix}
    \begin{pmatrix}
        J e^{i\theta} & \Delta e^{-i\varphi} \\ \Delta e^{i\varphi} & Je^{-i\theta} 
    \end{pmatrix}
    \begin{pmatrix}\hat a_{n+1}^\dag \\ \hat a_{n+1}\end{pmatrix},\label{eq:boson_KitaevXY}
\end{equation}
where $\hat{a}^\dagger_n$ ($\hat{a}_n$) is the bosonic creation(annihilation) operator at site $n$. 
$J$ is the tunneling amplitude with a phase $\theta$. 
$\Delta$ is the amplitude of the pair creation/annihilation and $\varphi$ is the corresponding phase. 
$\mu$ is the chemical potential. 
When $\mu=0$, this model was theoretically studied in~\cite{Clerk2018} and non-Hermitian dynamics were predicted to arise in a Hermitian system. 
This model was realized in two recent experiments~\cite{Slim2024, Clerk2024}. 

Though bosonic Kitaev model in Eq.~(\ref{eq:boson_KitaevXY}) appears similar to any other lattice models defined in a flat space, two hyperbolic surfaces are hidden there. 
To see this fact, we first consider $\mu=0$. 
Following the approach taken in~\cite{Clerk2018}, we express the bosonic operators as $\hat{a}_n=(\hat{X}_n+i\hat{P}_n)/\sqrt{2}$,  $\hat{a}^\dagger_n=(\hat{X}_n-i\hat{P_n})/\sqrt{2}$. 
Furthermore, since $\varphi$ can be gauged away, we set $\varphi=\pi/2$ for convenience. 
When $\Delta<J\cos(\theta)$, a unitary transformation maps Eq.~(\ref{eq:boson_KitaevXY}) to a simple tight-binding model~\cite{Clerk2018}.  Nontrivial physics arises in the regime where $\Delta>J\cos(\theta)$. 
A unitary transformation exists such that $\theta\rightarrow \pi/2$ and $\Delta\rightarrow \sqrt{\Delta^2-J^2\cos^2(\theta)},\,J\to J\sin(\theta)$. 
As such, it is sufficient to consider $\theta=\varphi=\pi/2$ in this regime.  
The equations of motion $d\hat X_n/d\tau=i[\hat H_b,\hat X_n]$ and $d\hat P_n/d\tau=i[\hat H_b,\hat P_n]$
are written as  
\begin{equation}
    \begin{split}
        \dot X_n=& (J+\Delta)X_{n-1} + (-J+\Delta)X_{n+1},\\
        \dot P_n=& (J-\Delta)P_{n-1} + (-J-\Delta)P_{n+1}.
    \end{split}\label{eq:KitaevHN}
\end{equation}
We have dropped the hats by taking the expectation values of operators. 
Each equation in Eq.~(\ref{eq:KitaevHN}) can be cast in the form of Hatano-Nelson chain, 
\begin{equation}
     \dot{\psi}_n=-\sum_m\mathcal{M}_{m,n} \psi_{m}, 
\end{equation}
where $\mathcal{M}_{m,n}$ is the matrix element of a matrix $\mathcal{M}$, and
\begin{equation}
     \mathcal{M}_{m,n}=t_R\delta_{m,n-1} + t_L \delta_{m,n+1}.\label{eq:HNmodel}
\end{equation}
$\psi_n$ represents $X_n$ or $P_n$ and $t_R(t_L)$ is chiral tunneling to the right(left). 

When $t_Lt_R>0$, eigenvalues of $\mathcal{M}$ in Eq.~(\ref{eq:HNmodel}) is real and bounded between $-2\sqrt{t_Lt_R}$ and $2\sqrt{t_Lt_R}$. 
A vanishing $t_Lt_R$ defines the so-called exceptional point, across which the eigenvalues become purely imaginary and bounded between $-2i\sqrt{|t_Lt_R|}$ and $2i\sqrt{|t_Lt_R|}$. 
Under this situation ($t_Lt_R<0$), Eq.~(\ref{eq:HNmodel}) can be rewritten in the form of a Schr\"odinger equation by defining $\psi_n=i^n\phi_n$, 
\begin{equation}
     i\dot{\phi}_n=-t_R \phi_{n-1} +t_L \phi_{n+1}. \label{eq:HNmodel2}
\end{equation}
Note that $t_Lt_R<0$, the two tunneling amplitudes have the same sign in Eq.~(\ref{eq:HNmodel2}). 
Alternatively, we may use $\psi_n=(-i)^n\phi_n$ and the right-hand side of Eq.~(\ref{eq:HNmodel2}) then has an extra minus sign. 
In the presence of an extra $i$ on its left-hand side, the imaginary eigenvalues of $\mathcal{M}$ become the real eigenenergy of the Schr\"odinger equation in Eq.~(\ref{eq:HNmodel2}).  

Adopting the recently discovered duality between non-Hermiticity and curved spaces~\cite{Lv2022}, the continuous limit of Eq.~(\ref{eq:HNmodel2}) near the band bottom ($-2\sqrt{|t_Lt_R}|$) can be written as a 1D non-relativistic Schr\"odinger equation with an imaginary vector potential,
\begin{equation}
   i\dot{\phi}(s)=\bigg[-\frac{\hbar^2}{2m}(\partial_s-A)^2+\Gamma\bigg]\phi(s),\label{eq:schro}
\end{equation}
where $\phi(s=nd)\equiv \phi_n$, $m=\hbar^2/(2\sqrt{|t_Lt_R|}d^2)$, $A=\ln(|t_R/t_L|)/(2d)$, $\Gamma=-2\sqrt{|t_Lt_R}|$ and $d$ is the lattice spacing. 
Near the band top ($2\sqrt{|t_Lt_R|}$), $m$ and $\Gamma$ change sign.  
Eq.~(\ref{eq:schro}) is the projection of a non-relativistic Schr\"odinger equation on a two-dimensional hyperbolic surface with constant negative curvature $-\kappa=-\ln^2(|t_R/t_L|)/d^2$,
\begin{equation}
   i\dot{\Phi}(s,x) = \bigg[{-\frac{\hbar^2}{2m}\frac{1}{\sqrt{g}}\partial_\mu g^{\mu\nu}\sqrt{g}\partial_\nu+\Gamma-
   \frac{\hbar^2\kappa}{8m}}\bigg]\Phi(s,x).\label{eq:hyperbolic}
\end{equation}
The line element on a hyperbolic surface is written as 
\begin{equation}
    dl^2=g^{\mu\nu}dx^\mu dx^\nu=ds^2+e^{-2\sqrt{\kappa} s}dx^2,\label{eq:metric}
\end{equation}
and the Laplacian operator is replaced by the Laplace-Beltrami operator~\cite{Wald1995,Mukhanov2007}.
Consider the periodic boundary condition in the $x$ direction, $\Phi(s,x)=e^{ik_xx}\phi(s)$. When $k_x=0$, Eq.~(\ref{eq:hyperbolic}) reduces to Eq.~(\ref{eq:schro}).

The embedding of a hyperbolic surface in a three-dimensional Euclidean space appears to be a funnel, as shown in Fig~\ref{fig:Kitaev} (b). 
All eigenstates are thus localized at the funneling mouth with the smallest circumference in the azimuthal direction. 
It is worth mentioning that we should consider real part of the wavefunction here, as both $X_n$ and $P_n$ are real and Eq.~(\ref{eq:KitaevHN}) is linear.
As such, the values of $X_n$ and $P_n$ oscillate periodically in the time domain. 
In addition to these eigenmodes, we could also consider a wavepacket traveling on a hyperbolic surface. 
It gets amplified (suppressed) when moving toward (away from) the mouth of the funnel as required by the conservation of the total probability~\cite{Lv2022}. 
Getting closer (further away from) the funneling mouth, the circumference in the azimuthal direction becomes smaller (larger) such that the amplitude of the wavefunction has to increase (decrease). 
As this funneling effect on a hyperbolic surface is unified with the non-Hermitian skin effect by the same underlying geometric effect, we thus simply dubbed it skin effect.  
Since the mouths of the two hyperbolic surfaces in bosonic Kitaev model are located at opposite edges, we see that it is the finite curvatures that push excitations in the $X$- and $P$-chains to propagate towards opposite directions. 
This unravels the geometric roots of the chiral transport discovered in bosonic Kitaev model. 

The hidden hyperbolic surfaces can also be understood from a discrete tree graph. 
Fig.~\ref{fig:Kitaev}(b) shows a simple example of a tree graph in which each node is connected to two nodes in the next layer, i.e., $q=2$. 
The Schr\"odinger equation on this tree graph is written as 
\begin{equation}
    i\dot{\phi}_{n,m}= -t (\phi_{n-1,[[\frac{m+1}{2}]]} + \phi_{n+1,2m} + \phi_{n+1,2m-1}) .\label{eq:tree_Schrodinger}
\end{equation}
where $[[x]]$ denotes the integer part of $x$.
Whereas this model is Hermitian and there is no chiral coupling between any pair of nodes, non-Hermiticity and chiral couplings arise once we consider a uniform distribution in the $m$-direction, i.e., {$\phi_{n,m}\equiv\Phi_n$} is independent of $m$. 
Substituting this expression into Eq.~(\ref{eq:tree_Schrodinger}), we immediately see that the equation of motion satisfied by $\Phi_n$ is given by Eq.~(\ref{eq:HNmodel}) where 
\begin{equation}
    t_L=2t,\,\,\,\,  t_R=t
\end{equation}
For a generic $q$, the above equation is modified to 
\begin{equation}
    t_L=qt,\,\,\,\, t_R=t.
\end{equation}
A tree graph, also known as the Bethe lattice, represents a particular tiling of a hyperbolic surface $(p=\infty,q)$, where $p$ and $q$ denote the number of edges of the polygon and the number of edges connected to each node~\cite{Mosseri1982}. 
We thus see that a bosonic Kitaev chain includes two tree graphs whose roots are located at opposite edges of the system. 
This is the discretized version of the previously discussed hyperbolic surfaces.  

Whereas we focused on $t_Lt_R<0$ in the previous discussions, the underlying hyperbolic surfaces also exist when $t_Lt_R>0$. 
For instance, around the band top ($2\sqrt{t_Lt_R}$), the continuum limit of Eq.~(\ref{eq:HNmodel}) becomes
\begin{equation}
   \dot{\bar \psi}(s)=[+D(\partial_s-A)^2-\Gamma]\bar \psi(s),\label{eq:dif}
\end{equation}
where $D=\sqrt{t_Lt_R}d^2$ is the diffusion constant, $\bar \psi(s)=e^{i\pi s/d} \psi(s)$. 
Near the band bottom ($-2\sqrt{t_Lt_R}$), $D$ and $\Gamma$ change sign, corresponding to anti-diffusion. 
The same $A$ as that in Eq.~(\ref{eq:schro}) provides a drift term and $\Gamma$ now becomes the reaction constant. 
Using a similar argument as before, Eq.~(\ref{eq:dif}) is a projection of a diffusion-reaction equation on a hyperbolic surface with a constant curvature,
\begin{equation}
    \dot{\Psi}(s,x)=\bigg[D\frac{1}{\sqrt{g}}\partial_\mu g^{\mu\nu}\sqrt{g}\partial_\nu-\Gamma+
   \frac{\kappa D}{4}\bigg]\Psi(s,x).
\end{equation}
The line element is still given by Eq.~(\ref{eq:metric}) with the same $\kappa$. 
In other words, crossing the exceptional point signifies the change from a Schr\"odinger equation to a diffusion-reaction equation on the same hyperbolic surface. 

\begin{figure}
    \includegraphics[width=0.48\textwidth]{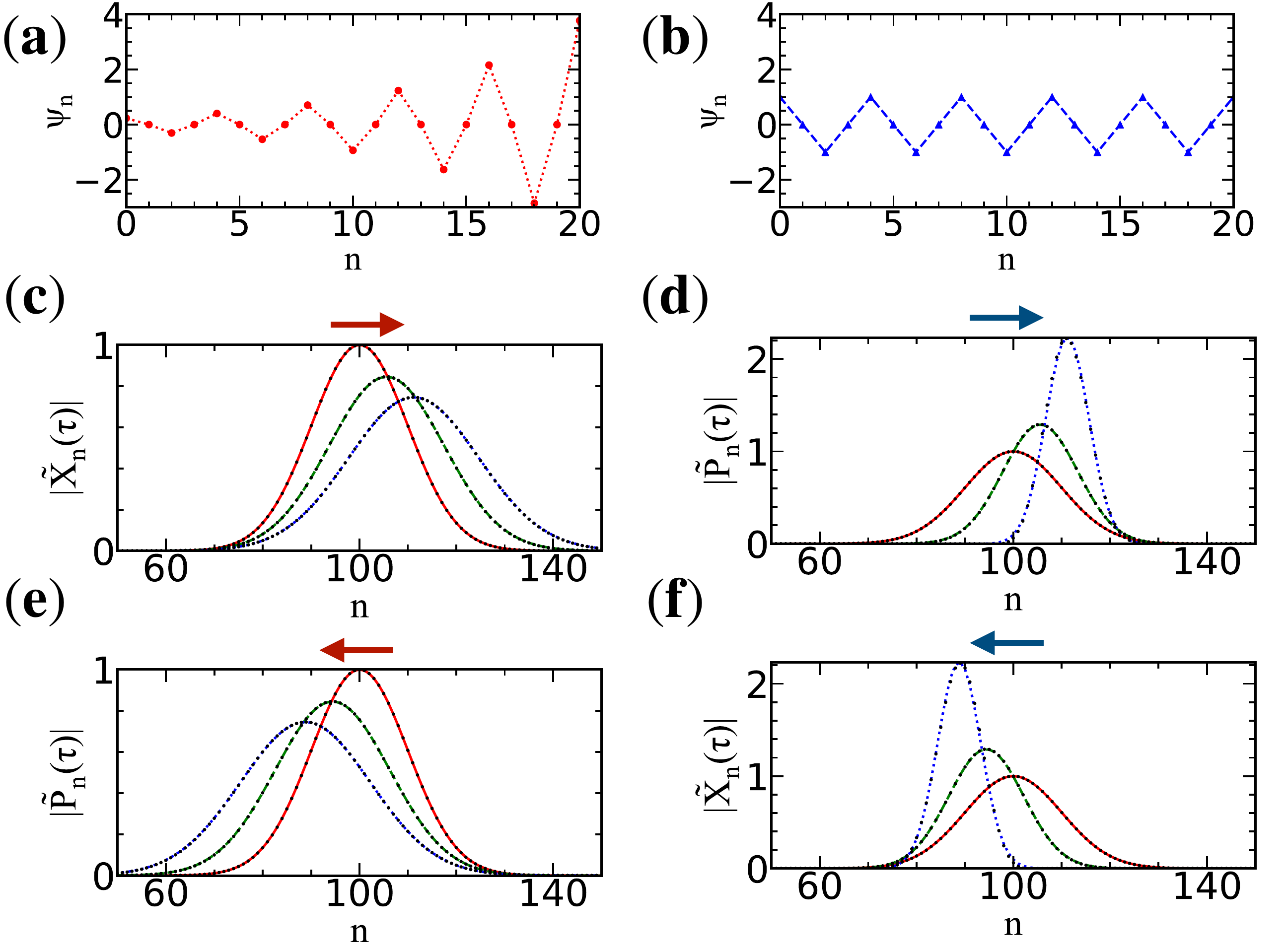}
    \caption{
    (a,b) Stable mode of Eq.~(\ref{eq:HNmodel}) for $t_Lt_R>0$. (a) $\sqrt{t_R/t_L} = 1.15$ 
    (b) $\sqrt{t_R/t_L} = 1$.
    (c, e) Chiral diffusion and (d, f) anti-diffusion of a Gaussian wavepacket evolved by Eq.~(\ref{eq:KitaevHN}) for $\Delta > J$ and $\sqrt{(\Delta+J)/(\Delta-J)}=1.15$. 
    Numerical results at $\sqrt{\Delta^2 - J^2}\tau=0,20,40$ are denoted by red solid, green dashed, and blue dotted curves. 
    Dots are obtained from the continuous effective theory. 
    The Gaussian wavepacket is prepared in $X$ near the band top (c), band bottom (f), and in $P$ near the band top (e), band bottom (d), respectively.
    The global gain or loss has been subtracted by considering $\tilde{X}_n=e^{\mp\gamma\tau}X_n$ and  $\tilde{P}_n=e^{\mp\gamma\tau}P_n$ near the band top or bottom.
    }\label{fig:diffusion}
\end{figure}

Unlike a finite $\Gamma$ in Eq.~(\ref{eq:hyperbolic}) that only leads to an unimportant phase factor, $\Gamma$ in Eq.~(\ref{eq:dif}) generically leads to either unbounded grow or loss. 
Still, interesting results of stationary solutions arise from the interplay of diffusion and reaction. 
In the lattice model, this can be seen that $\psi^{[0]}_n= i^n(t_R/t_L)^{n/2}$ or $\psi^{[1]}_n=(-i)^n(t_R/t_L)^{n/2}$ leads to a vanishing right-hand side of Eq.~(\ref{eq:HNmodel}). 
As such, any superposition of these two independent solutions leads to a stationary state. 
Real stationary states for $\sqrt{t_R/t_L}\neq 1$ and $\sqrt{t_R/t_L}= 1$ are plotted in Fig.~\ref{fig:diffusion} (a) and (b). 
A finite $\kappa$ makes the stationary state exponentially localized near the boundary.
Note that both $\psi^{[0]}$ and $\psi^{[1]}$ have a wavevector $K_{\pm}=\pm\pi/(2d) - i\ln(|t_R/t_L|)/(2d)$, the effective theory for $\psi$ with small excitations on top of these stationary states need to be formulated around $K_{\pm}$, 
\begin{equation}
    \dot \psi(s)=\mp iv_s(\partial_s-iK_\pm)\psi(s),\label{eq:eff_stationary}
\end{equation}
where $\psi(s=nd)=\psi_n$, $v_s=2\sqrt{|t_Rt_L|}d$ (Supplementary Matrials).

In addition to the stationary states, we also consider the generic solution to Eq.~(\ref{eq:dif}). 
Whereas a finite $\gamma\equiv -\Gamma+\kappa D/4$ leads to a global gain (or loss), once we deduct such a gain (loss) from $\psi(s,\tau)$ by defining $\tilde{\psi}(s,\tau)=e^{\mp\gamma \tau}\psi(s,\tau)$, $\tilde{\psi}(s,\tau)$ is governed by a diffusion equation in the absence of reaction. 
However, distinct from the ordinary diffusion equation, a finite $A$ here now drifts the excitation toward a unique direction. 
Furthermore, since the drift constant $A$ has opposite signs in the equation for $X_n$ and $P_n$, excitations move toward opposite directions in the $X$- and $P$-chain. 
This chiral diffusion-reaction, which occurs when $t_Lt_R>0$ is the counterpart of the previously discussed chiral quantum transport when $t_Lt_R<0$. 
As shown by Fig.~\ref{fig:diffusion}(c-f), we have compared the results of Eq.~(\ref{eq:KitaevHN}) for $\tilde\psi$ and the effective theory based on Eq.~(\ref{eq:dif}) when the initial wavepacket is prepared at the band top or bottom. 
As expected, these results agree with each other well. 
In the diffusion case ($D>0$), the amplitude decreases, as the width of the wavepacket increases fast and overcomes the amplification due to finite $\kappa$ when the wavepacket moves to the small funneling mouth. 
In the anti-diffusion case ($D<0$), the width of the wavepacket decreases and further amplifies the amplitude.

We now turn to a finite $\mu$. 
Whereas $\mu$ vanishes for photons considered in~\cite{Clerk2018}, it is important to consider a generic bosonic system that has a finite chemical potential. 
In fact, taking into account a finite $\mu$ unfolds much richer physics. 
The previously discussed two Hatano-Nelson chains are coupled and Eq.~(\ref{eq:KitaevHN})now becomes 
\begin{equation}
    \begin{split}
        \dot X_n=&+\mu P_n+(J+\Delta)X_{n-1}+(-J+\Delta)X_{n+1}\\
        \dot P_n=&-\mu X_n+(J-\Delta)P_{n-1}+(-J-\Delta)P_{n+1}.
    \end{split}\label{eq:KitaevHN_coupled}. 
\end{equation}
A finite $\mu$ leads to the
critical skin effect, where a small inter-chain coupling produces significant changes in physical properties of the system~\cite{Li2020,Liu2020,Yokomizo2021,Rafi2022_1,Rafi2022_2,Qin2023}. 
In particular, the larger the size of the system is, the more drastic the change is. 
While physicists attempted to understand such critical skin effects using the generalized Brillouin zone~\cite{Yao2018,Yokomizo2019,Yang2020}, there still lacks a transparent physical interpretation of this remarkable phenomenon. 
Here, the underlying physics of the critical skin effect becomes immediately clear once we note the hidden curved spaces.   

To be explicit, the continuum limit of Eq.~(\ref{eq:KitaevHN_coupled}) at the band bottom is written as 
\begin{equation}
    \begin{split}
        i\dot \psi_X(s)=&+i\mu \psi_P(s)-\frac{\hbar^2}{2m}(\partial_s-A_-)^2\psi_X(s)\\
        i\dot \psi_P(s)=&-i\mu \psi_X(s)-\frac{\hbar^2}{2m}(\partial_s+A_+)^2\psi_P(s).
    \end{split},\label{eq:coupledhyper}
\end{equation}
where $A_\pm=\ln((J+\Delta)/(J-\Delta))/(2d)\pm i\pi/(2d)$, $\psi_X(s=nd)\equiv X_n$ and $\psi_P(s=nd)\equiv P_n$. 
When $\mu$ vanishes, Eq.~(\ref{eq:coupledhyper}) describes two decoupled hyperbolic surfaces, as previously discussed. The energy spectrum is doubly degenerate, as these two hyperbolic surfaces are simply mirror images of each other. Due to the skin effect, the two degenerate eigenstates are located at opposite edges of the system. 

Now, consider a small $\mu$ that is much less than the energy separations in the unperturbed energy spectrum. 
A perturbative approach can be used and the effective theory is given by a two-by-two matrix $M$. 
Here, we need to remind ourselves that each matrix element in a curved space needs to include the metric tensor. 
For instance, the inner products of two states in a curved space whose metric tensor is given by Eq.~(\ref{eq:metric}) is written as 
\begin{equation}
  \bra{\Phi}\ket{\Phi'}=\int dxds\sqrt{g}\Phi^{*}(x,s)\Phi'(x,s),  
\end{equation}
where $g=e^{-2\sqrt{\kappa}s}$ is the determinant of the metric tensor. 
If a periodic boundary condition is chosen in the $x$-direction and the momentum is set as 0, the above equation reduces to 
\begin{equation}
  \bra{\phi}\ket{\phi'}=\int \sqrt{g}ds\phi^*(s)\phi'(s)\label{eq:orthnormal},
\end{equation}
where $\phi(s)=\Phi(x,s)$ ($\phi'(s)=\Phi'(x,s)$).
In the language of non-Hermitian physics, the extra piece was referred to as the metric operator in the literature, but its physical meaning remained unclear until the duality between non-Hermiticity and curved spaces was discovered. 
This duality shows that such a metric operator does have a geometric meaning as it arises from the underlying curved space. 
Here, a finite $\sqrt{g}$ is precisely the source of the critical skin effect. 

The geometric origin of the critical skin effect can also be seen in the discrete lattice model. 
The two degenerate ground states are written as 
\begin{equation}
    \psi_X(n)=\sqrt{\frac{2}{L+1}}\bigg(\frac{J+\Delta}{J-\Delta}\bigg)^{\frac{n-L}{2}}\sin(\frac{\pi}{L+1}n)i^n, 
\end{equation}

\begin{equation}
    \psi_P(n)=\sqrt{\frac{2}{L+1}}\bigg(\frac{J-\Delta}{J+\Delta}\bigg)^{\frac{n-1}{2}}\sin(\frac{\pi}{L+1}n)i^n.
\end{equation}
In conventional perturbative approaches in a flat space, the off-diagonal matrix elements are simply $M_{12}=\sum_n \psi^*_P(n)\psi_X(n)$ and  $M_{21}=\sum_n \psi^*_X(n)\psi_P(n)$. 
For the lattice model we are considering, we just use the discrete version of Eq.~(\ref{eq:orthnormal}). 
In the discrete model, $\sqrt{g}$ becomes the number of lattice sites in each layer of the tree graph. 
As such, both $M_{12}$ and $M_{21}$ depend on the corresponding metric tensor of one of two Hatano-Nelson chains. 
$M_{12}$ corresponds to the annihilation in the $X$-chain and the creation in the $P$-chain, i.e., a particle tunnels from the $X$-chain to the $P$-chain, this matrix element should include the metric $\sqrt{g}$ of the $P$-chain, $[(J+\Delta)/(J-\Delta)]^{n-1}$
\begin{equation}
M_{12} = \sum_{n=1}^L \bigg(\frac{J+\Delta}{J-\Delta}\bigg)^{n-1} \psi^*_P(n)\psi_X(n)~\label{eq:M12}.
\end{equation}
Similarly, $M_{21}$ must include the metric of the $X$-chain, $[(J+\Delta)/(J-\Delta)]^{L-n}$,
\begin{equation}
M_{21} = \sum_{n=1}^L \bigg(\frac{J+\Delta}{J-\Delta}\bigg)^{L-n} \psi^*_X(n)\psi_P(n)~\label{eq:M21} .
\end{equation}

\begin{figure}
    \includegraphics[width=0.48\textwidth]{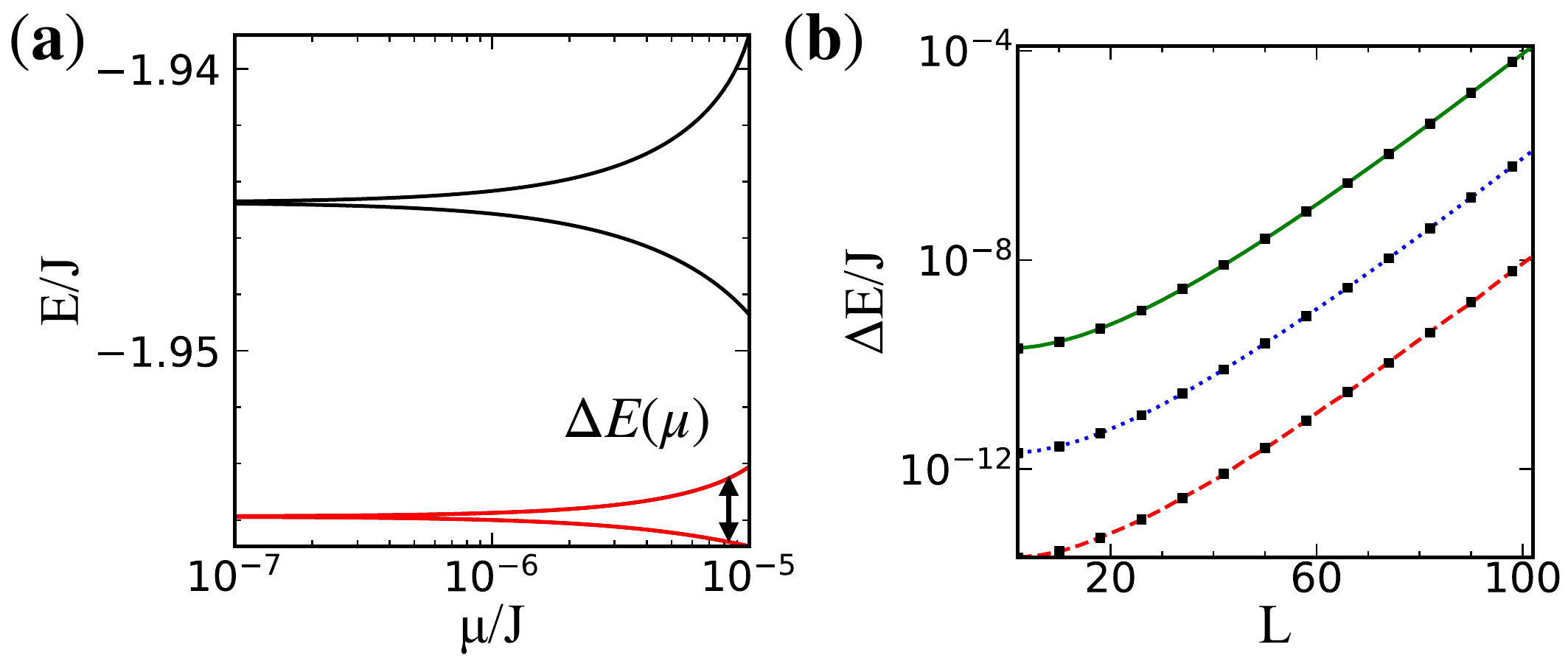}
    \caption{
        (a) Eigenenergies as a function of $\mu$ of bosonic Kitaev model. 
        Results of $L=50$, $\Delta/J=0.2$ are shown. $\Delta E$ denotes the energy splitting between the two degenerate ground states at $\mu=0$ once $\mu$ becomes finite.  
        (b) The scaling of $\Delta E$ with the number of lattice site $L$. The red dashed, blue dotted, and green solid curves correspond to $\mu/J=10^{-14}, 10^{-12}, 10^{-10}$, respectively. 
        The black boxes are analytical perturbation results in the curved space.}\label{fig:gap_sensitivity}
\end{figure}
A straightforward calculation shows that both off-diagonal matrix elements grow exponentially with increasing the size of the system, and the 2 by 2 matrix can be written as 
\begin{equation}
 M = \frac{4\mu(\chi^2+1) \sin^2(\frac{\pi }{L+1}) \chi^{2} \sinh((L+1)\ln(\chi))}{(L+1) (\chi^2 -1) |\chi^2 -e^{\frac{2 i \pi }{L+1}}|^2}\sigma_y
\end{equation}
where $(J+\Delta)/(J-\Delta)=\chi^2$, $\sigma_y$ is the Pauli-Y matrix. When $L\rightarrow \infty$, $M_{12}\rightarrow \chi^L/L^3$. As such, the energy splitting $\sim \mu\chi^L/L^3$. 
This means that even a tiny $\mu$ could have significant effects in a large system. 
In other words, a bosonic Kitaev chain can be used as a quantum sensor to measure the chemical potential whose sensitivity grows exponentially with the size of the system.
Fig.~\ref{fig:gap_sensitivity} depicts the energy gap opening of the ground band for different $\mu$ at a finite $\Delta$ and its scaling with the number of lattice sites $L$. 
We see that the exact numerical results agree with the degenerate perturbation theory in the curved space. 

As seen from previous discussions, the amplification of the effects of a small $\mu$ has a deep geometric origin that can be traced back to Fig.~\ref{fig:Kitaev}. 
A finite $\mu$ couples two funnels, whose funneling mouths are located at opposite sides. 
When $\Delta=0$ or equivalently, $t_L=t_R$, these two hyperbolic surfaces reduce to two cylinders with vanishing curvatures and such an exponential amplification disappears. 
As such, a bosonic Kitaev chain delivers a new class of geometry-based quantum sensors whose supersensitivity is entirely induced by a finite curvature. 
Whereas it may be challenging to realize couplings between two realistic curved spaces as shown in Fig.~\ref{fig:Kitaev}, a bosonic Kitaev chain with a finite $\mu$ naturally encodes such couplings between the hidden two hyperbolic surfaces.  

We have shown that hyperbolic surfaces are hidden in bosonic Kitaev model. 
The interplay between the tunneling and the pair correlation/annihilation produces finite curvatures and intriguing chiral transport. 
Furthermore, quantum sensors could be generated as the finite curvatures make bosonic Kitaev chains supersensitive to the chemical potential. 
It is promising that our theoretical predictions can be tested in laboratories in the foreseeable future. Whereas we focus on bosonic Kitaev model with uniform couplings, which lead to hyperbolic surfaces with constant curvature, it is expected that our results can be generalized to other models in which local curvatures may be tailored such that even richer phenomena may arise from curved spaces hidden in quantum models. 

{\bf Acknowledgement}  We thank Thomas Bilitewski, Rebekah Hermsmeier, Timur Tscherbul, and Ana Maria Rey for helpful discussions. This work is supported by The U.S. Department of Energy, Office of Science through the Quantum Science Center (QSC), a National Quantum Information Science Research Center, Army Research Office under award number FA9550-23-1-0491, and Air Force Office of Scientific Research under award number FA9550-20-1-0221.

\onecolumngrid
\newpage
\vspace{0.4in}
\centerline{\bf\large Supplementary Materials for ``Hidden curved spaces in bosonic Kitaev models"}
\setcounter{equation}{0}
\setcounter{figure}{0}
\setcounter{table}{0}
\makeatletter
\renewcommand{\theequation}{S\arabic{equation}}
\renewcommand{\thefigure}{S\arabic{figure}}
\renewcommand{\thetable}{S\arabic{table}}
\vspace{0.2in}

\section{Effective theory at other $K$}
In the main text, we explore the effective theory for $t_Lt_R>0$ at ${\rm Re}(K)=0,\pi/d,\pm \pi/(2d)$. 
Effective theories similar to Eq.~(14) with a reaction term exist at other $K$. 
To derive the effective theory near a general $K$, we first define $\psi_n(\tau) = e^{iKnd}\phi_n(\tau)$.
The equation of motion for $\phi_n$ is 
\begin{equation}
    \dot\phi_n(\tau) = - t_R e^{-iKd}\phi_{n-1}(\tau) - t_L e^{iKd}\phi_{n+1}(\tau).
\end{equation}
Let $\phi(s=nd)\equiv \phi_n$ and apply Taylor expansion to the second order $O(d^2)$, we find 
\begin{equation}
    \dot\phi(s,\tau) = - \bigg[t_R e^{-iKd} + t_L e^{iKd} + (-t_R e^{-iKd} + t_L e^{iKd})d\partial_s + \frac{1}{2}(t_R e^{-iKd} + t_L e^{iKd})d^2\partial_s^2\bigg]\phi(s,\tau).
\end{equation}
Using $\psi(s=nd)\equiv \psi_n$ and $\phi(s,\tau) = e^{-iKs}\psi(s,\tau)$, we obtain the continuous effective theory of $\psi(s,\tau)$,
\begin{equation}
    \dot\psi(s,\tau) = - \bigg[t_R e^{-iKd} + t_L e^{iKd} + (-t_R e^{-iKd} + t_L e^{iKd})d(\partial_s-iK) + \frac{1}{2}(t_R e^{-iKd} + t_L e^{iKd})d^2(\partial_s-iK)^2\bigg]\psi(s,\tau).
\end{equation}
Put $K=K_0-i\ln(t_R/t_L)/(2d)$ into the equation above,
\begin{equation}
    \dot\psi(s,\tau)=[-i\sin(K_0d)v_s(\partial_s-iK)+\cos(K_0d)\Gamma -D\cos(K_0d)(\partial_s-iK)^2]\psi(s,\tau).
\end{equation}
It becomes Eq.~(14) in the main text at $K_0=\pm \pi/(2d)$.
When $K_0\neq \pm \pi/(2d)$, $\cos(K_0d)\Gamma$ dominates. We need to subtract the unbounded growth/decay to observe the dynamics generated by the kinetic terms.
When $K_0=0,\pi/d$, $v_s$ vanishes. We obtain the diffusion/anti-diffusion equation with a reaction term,
\begin{equation}
    \dot\psi(s,\tau)=\pm[\Gamma -D(\partial_s-iK)^2]\psi(s,\tau).
\end{equation}

\end{document}